\documentclass{emulateapj-rtx4}
\newcommand{\hi}{\protect\ion{H}{1}}
\newcommand{\hii}{\protect\ion{H}{2}}

\newcommand{\kms}{~km\,s$^{-1}$}

\usepackage{lscape}

\begin{document}
\title{NGC~4656UV: A UV-selected Tidal Dwarf Galaxy Candidate\footnote{Based in part on observations made with the NASA Galaxy Evolution Explorer.
GALEX is operated for NASA by the California Institute of Technology under NASA contract NAS5-98034.}}
\author{Andrew Schechtman-Rook\altaffilmark{a}}
\author{Kelley M. Hess\altaffilmark{b,a}}

\altaffiltext{a}{Department of Astronomy, University of Wisconsin -- Madison, 475 N. Charter St., Madison, WI 53706}
\email{andrew@astro.wisc.edu}
\altaffiltext{b}{University of Cape Town, Department of Astronomy, Private Bag X3, Rondebosch 7701}
\email{hess@ast.uct.ac.za}
\shorttitle{NGC~4656UV}

\begin{abstract}
We report the discovery of a UV-bright tidal dwarf galaxy candidate in the NGC 4631/4656 galaxy group, which we designate NGC~4656UV. Using survey and archival data spanning from 1.4 GHz to the ultraviolet we investigate the gas kinematics and stellar properties of this system. The \hi\ morphologies of NGC~4656UV and its parent galaxy NGC~4656 are extremely disturbed, with significant amounts of counterrotating and extraplanar gas. From UV-FIR photometry, computed using a new method to correct for surface gradients on faint objects, we find that NGC~4656UV has no significant dust opacity and a blue spectral energy distribution.  We compute a star formation rate of 0.027 $M_{\odot}$ yr$^{-1}$ from the FUV flux and measure a total \hi\ mass of 3.8$\times 10^{8}\ M_{\odot}$ for the object.  Evolutionary synthesis modeling indicates that NGC~4656UV is a low metallicity system whose only major burst of star formation occurred within the last $\sim260-290$ Myr.  The age of the stellar population is consistent with a rough timescale for a recent tidal interaction between NGC~4656 and NGC~4631, although we discuss the true nature of the object--whether it is tidal or pre-existing in origin--in the context of its metallicity being a factor of ten lower than its parent galaxy.  We estimate that NGC~4656UV is either marginally bound {\it or} unbound.  If bound, it contains relatively low amounts of dark matter. The abundance of archival data allows for a deeper investigation into this dynamic system than is currently possible for most TDG candidates.
\end{abstract}

\keywords{galaxies: individual (NGC~4656) -- galaxies: dwarf -- galaxies: interactions -- galaxies: kinematics and dynamics -- galaxies: structure}

\section{Introduction}
Systems of interacting galaxies have long been known to contain some very irregular structures such as rings, bridges, plumes, and tidal tails (see \citealt{Schweizer86} for a review).  In recent years, tidal dwarf galaxies (TDGs) have also been recognized as byproducts of collisions between massive galaxies, identified by regions of active star formation and reservoirs of cold gas (see \citealt{Duc99,Duc11} for a review). Simulations show that they consist of baryonic material stripped from the outer disks of interacting gas-rich 'parent' galaxies \citep{Hibbard95, Duc00}.  TDGs are believed to form stars through gas instabilities and collapsing \hi\ clouds, and contain little to no dark matter \citep{Barnes92,Elmegreen93}.  In general, TDGs appear to fall into one of two categories based on their stellar population: those dominated by old stars ripped from their parent galaxies, and those that consist almost exclusively of new stars forming from the \hi\ gas that was stripped during the interaction \citep{Duc99,Duc00,Braine01}. 

Tidal features are interesting for several reasons. They reveal the history of galaxy interactions \citep{Hibbard94} and they may be used to trace the underlying dark matter distribution of the group or cluster (e.g. \citealt{Mihos98}).  The ultimate fate of TDGs is unknown, but due to their dearth of dark matter they should be minimally bound by self-gravity and therefore easily disrupted by further galaxy encounters.  They provide a potential source for metal enrichment of the intergalactic medium \citep{Wiersma10} as the material they contribute will have a similar metallicity to their already-evolved parent galaxy \citep{Duc94,Duc99}. They are also a proposed origin of intracluster light \citep{Gregg98,Mihos05}.  Finally, TDGs provide a laboratory for studies of star formation in low density environments which are not complicated by the dynamics of spiral galaxies \citep{Boquien09}.

Neutral hydrogen observations play a crucial role in identifying TDGs.  The \hi\ gas in late-type galaxies typically extends far beyond the stellar disk (e.g. \citealt{vanderKruit84}), so it is easily disturbed or removed in gravitational encounters, revealing the kinematics of an interaction. Numerical simulations have successfully reproduced the distribution of gaseous and stellar tidal debris in interacting systems (e.g. \citealt{Toomre72}).
A large fraction of this gas may fall back to the parent galaxy \citep{Hibbard94}, but some portion of it is pulled out with stars.  As a result, we observe gaseous debris among tidal streams, and TDGs typically contain their own reservoir of \hi\ gas out of which they are currently forming stars.  Instabilities in this gaseous component may be the driving factor in the formation of TDGs \citep{Barnes92,Elmegreen93,Duc01}.  The velocity dispersion of \hi\ may be used to estimate the dynamical mass and evaluate whether \hi\ density enhancements are gravitationally bound.  However, the \hi\ can be entangled in tidal material. CO gas, which forms in-situ, may provide a more accurate estimate of the TDG structure \citep{Braine01}.

Generally, TDGs are observed to be young systems which may be an indicator of their short lifetime \citep{Hibbard95} or may be due to the difficulty in identifying older TDGs \citep{Duc07}.   However, they are also generally observed to be chemically evolved: TDGs contain higher metallicities than normal dwarf galaxies, up to approximately $\frac{1}{3}$ solar, and they are abundant in molecular gas as measured by CO observations \citep{Braine01}.  Thus, metallicity estimates are used to confirm that TDG material is similar to that of its putative parent galaxy while population synthesis modeling is used to show that the stellar ages are less than the age of the tidal feature to which they belong (e.g. \citealt{Weilbacher00}).

How star formation proceeds in TDGs remains unknown and may provide valuable clues for how star formation occurs in general.  \citet{Boquien09} find evidence that the star formation properties in intergalactic star forming regions are similar to that in spiral galaxies. However, other observations suggest that star formation in low density environments, such as  in dwarf galaxies or the extreme outer disks of massive galaxies, may take place below classical star formation gas surface density thresholds put forth by \citet{Kennicutt98} and \citet{Martin01} (e.g. \citealt{Thilker07,Roychowdhury09}).  Further, scatter in the predicted star formation rates from UV, H$\alpha$, and 8.0 $\mu$m observations in these low density environments is not understood, but may be a function of age, metallicity, gas content, dust content, or the stellar initial mass function \citep{Thilker11}.  As dynamically simple systems with a range of metallicities, molecular gas, and dust content, dwarf and tidal dwarf galaxies may be good laboratories for shedding light on the mechanics of star formation.

We serendipitously discovered a UV-bright intergalactic star forming region, which was previously unrecognized or unacknowledged in optical observations, while examining GALEX images of the edge-on galaxy NGC~4656.  We consider the object (hereafter referred to as NGC~4656UV) a strong candidate for tidal dwarf galaxy status.

NGC~4656UV is a member of the loose (total number of members $\sim$6; \citealt{Garcia93}) galaxy group consisting primarily of two edge-on spiral galaxies, NGC~4656 and NGC~4631, which are suspected to have had a strong interaction in the past.  An \hi\ bridge discovered by \citet{Roberts68} extends between the two spirals suggesting gas has been stripped from NGC~4631 by its gravitational encounter with NGC~4656.   High resolution interferometric observations of the group resolve the bridge into multiple \hi\ ``fingers'' \citep{Weliachew78,Rand94}.  In particular, \cite{Rand94} notes that NGC~4656 contains evidence of a disturbed morphology resulting from a prior interaction, including non-circular motions and asymmetry. In the optical, previous attention has been paid to the sharp bend at the northeast edge of NGC~4656, a feature which gives the galaxy its distinct ``hockey stick'' appearance.  \citet{Stayton83} used photometry to determine that the bend is a feature of the galaxy resulting from an interaction, as opposed to a background galaxy (NGC~4657). 

An ultra-faint H$\alpha$ ``sheet'' has been detected between NGC~4656 and NGC~4631 \citep{Donahue95}, however NGC~4656UV is not detected in these observations.
\hi\ total intensity maps presented by \cite{Rand94} show a density enhancement at the location of NGC~4656UV, however they suggest it is part of an \hi\ ring surrounding NGC~4656 and do not identify the gas as being distinct from the parent galaxy. 

While nearly invisible in the Digitized Sky Survey and very close to detection limits in the red ({\it i,z}) Sloan Digital Sky Survey (SDSS, \citet{York00}) bands, GALEX observations easily reveal the TDG candidate.  NGC~4656UV is located just to the northeast of NGC~4656 at $12^h44^m14.794^s$ $+32^\circ16^m48.26^s$ (J2000) with a center-to-center separation of $\sim$11 kpc, assuming a distance of 7.2 Mpc \citep{Seth05}. NGC~4656 and NGC~4656UV are connected by a faint bridge of \hi\ gas and stars.  At this distance NGC~4656UV is one of the most nearby examples of vigorous star formation occurring in an extreme, low density environment. 

In this paper we present our investigation into the nature of this UV-bright tidal dwarf galaxy candidate through the accumulation of publicly available survey and archival data.  We make the case that it is most likely a tidal dwarf galaxy created from an encounter between NGC~4656 and NGC~4631, instead of a pre-existing, but highly rejuvenated, dwarf galaxy, or a tidal stream.  The paper is organized as follows: we describe the archival data and our post-processing methods in \S \ref{data}, we present our results in \S \ref{results} and discuss their implications in \S \ref{discussion}.  We end with concluding remarks in \S \ref{conclusion}.

\section{Data}
\label{data}
A plethora of survey and archival data exists on NGC~4656 and its companion candidate TDG across a broad wavelength regime, giving us nearly complete wavelength coverage from 2.01 keV to 21 cm. In addition to the discovery GALEX tiles, we obtained images from the SDSS, Two Micron All-Sky Survey (2MASS, \citet{Skrutskie06}), Infrared Array Camera (IRAC) and Multi-band Imaging Photometer (MIPS) on the {\it Spitzer Space Telescope}, and archival \hi\ observations from the Very Large Array (VLA). ROSAT X-ray imaging of NGC~4656 includes the TDG candidate, however there is no hard or soft X-ray emission from NGC~4656UV down to the detection limits of the observation ($\sim 9\times 10^{37}$ erg s$^{-1}$; Figure 8 of \citealt{Vogler96}).  

\subsection{HI 21 cm observations}
We retrieved archival VLA data from the online National Radio Astronomy Observatory Science Data Archive\footnote{https://archive.nrao.edu/archive/advquery.jsp}.  The observations consist of a single pointing in D-array taken on 18 December 1989 (Project Code AR0215).  It is centered on the optically bright portion of the edge-on spiral galaxy, NGC~4656 ($12^h43^m57.536^s$   $+32^{\circ}10'06.5''$, J2000). The primary beam of the VLA at L-band is half a degree in diameter, such that NGC~4656 and NGC~4656UV span nearly the full extent of the beam.  The galaxy was observed for approximately six hours on source with a velocity resolution of 10.3\kms.  We calibrated and imaged the data in AIPS using the standard procedures.  The resulting data cube has an RMS noise of 0.81 mJy beam$^{-1}$ channel$^{-1}$.  At 5$\sigma$, we achieved a mass sensitivity of $5.10\times10^5$ M$_{\odot}$ channel$^{-1}$ and a column density of $2.51\times10^{19}$ cm$^{-2}$ channel$^{-1}$.

The NGC~4631/NGC~4656 group of galaxies has been previously observed by \citet{Weliachew78} and by \citet{Rand94}, both with multiple pointings of the Westerbork Synthesis Radio Telescope (WSRT).  The archival VLA observations presented here are a single pointing focused on NGC~4656 and offer similar spectral and spatial resolution to the WSRT observations. However, the VLA observations provide an improvement of 1.7 and 1.4 times the sensitivity compared to the \citet{Weliachew78} and \citet{Rand94} WSRT observations, respectively.

\subsection{UV, optical, \& mid-IR imaging}
We visually inspected all UV, optical, and infrared images to ensure that they were free from defects and contained both the main galaxy and the TDG candidate object. Only about half of NGC~4656UV is on {\it Spitzer} IRAC 4.5 and 8.0 $\mu$m images, so we exclude those bands from photometry of the TDG candidate. To ensure complete and uniform coverage of the objects we downloaded image mosaics of the SDSS and 2MASS data using the Montage mosaicking software. 

An initial list of sources were extracted using the Source-Extractor package \citep{Bertin96}. Because SExtractor is not optimized for nearby extended sources, we hand-edited catalogs to remove spurious/overlapping ellipses and add masking regions over unblanked stars and background galaxies. To ensure uniform processing of the images, the hand-edited catalogs were homogenized for GALEX, SDSS, and MIPS bands. Due to the high number of contaminant point sources in the IRAC frames, additional masking was done for those bands. These extra sources were not blanked in the other images to maximize the accuracy of the photometry. In all cases we chose to mask sources where we were unsure if they were contaminants. 

\begin{figure}
\begin{centering}
\plotone{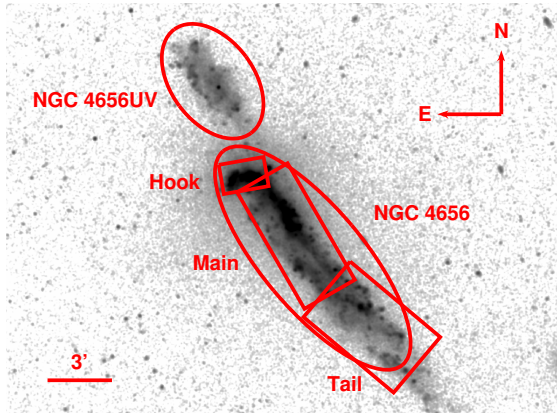}
\caption{GALEX FUV image of NGC~4656 and NGC~4656UV with the apertures used for photometry indicated. The smaller ellipse is for NGC~4656UV while the larger ellipse denotes NGC~4656. The boxes show the regions used for (from North-East to South-West) the Hook, Main, and Tail areas of NGC~4656.}
\label{fig:regions}
\end{centering}
\end{figure}

\begin{figure*}
\begin{centering}
\plotone{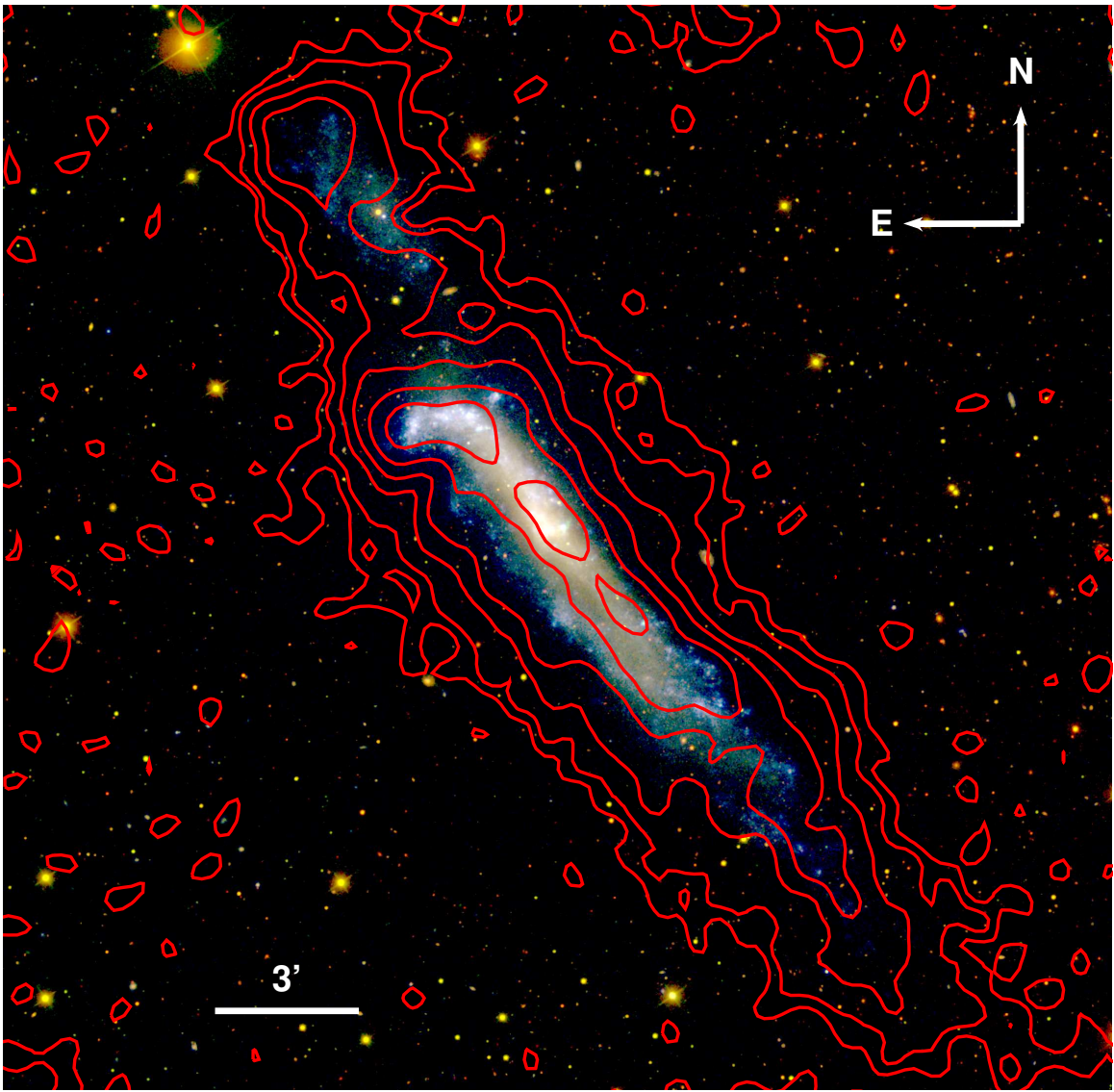}
\caption{False color RGB image of NGC~4656 and NGC~4656UV. Blue channel: GALEX FUV. Green channel: SDSS g. Red channel: SDSS i. The SDSS images have been smoothed to a resolution of $\sim$1''. Red contours are for \hi\ column density, drawn at 3, 8, 16, 32, 64, 128$\sigma$ where $\sigma = 2.61\times10^{19}$ cm$^{-2}$. While both sides of NGC~4656 are extremely blue, only NGC~4656UV has distinct separation between its \hi\ contours and those of its parent galaxy.}
\label{imagewcontours}
\end{centering}
\end{figure*}

We photometered both NGC~4656 and NGC~4656UV. We set elliptical apertures
around both objects, with sizes chosen to encompass all UV emission. We also
photometered three separate regions of NGC~4656, representing visually
distinct areas of the galaxy. On the northeast side of NGC~4656 (toward
NGC~4656UV) the `hook' is a significant 90$^\circ$ warp
seen in images from the UV to the far-infrared. This feature, part of which is
also known as NGC~4657, appears very dense and blue. We separate the
central portion of NGC~4656 as the `main' body. While there is no obvious
bulge in NGC~4656 to denote the center of the system, this region has both the reddest optical colors and the
highest surface-brightness of the galaxy. Finally, we photometer the extended
`tail' on the Southwest side of NGC~4656. The tail becomes very diffuse, and
there is a color gradient along the major axis of the galaxy with the
outermost regions of the tail becoming bluer. 
 The exact shape and extent of the regions are shown in Figure
 \ref{fig:regions} and detailed in Table \ref{tab:regions}, while a color
 image of the system is shown in Figure \ref{imagewcontours}. We note that in
 Figure \ref{fig:regions} UV emission is visible further Southwest than the
 end of the tail region; this region is off the edge of the IRAC 3.6 and 5.8
 $\mu$m images, so we do not include this area in all bands to preserve the
 colors of the spectral energy distributions (SEDs). For all bands and regions we adopt a detection threshold of S/N = 10.

\begin{deluxetable*}{c c c c c c}
\tabletypesize{\footnotesize}
\tablewidth{0pt}
\tablecaption{Apertures}
\tablehead{\colhead{Name} & \colhead{Region Type} & \colhead{RA} & \colhead{Declination} & \colhead{Dimensions$^{a}$} & \colhead{Position Angle} \\
& & \colhead{(J2000)} & \colhead{(J2000)} & \colhead{(arcseconds)} & \colhead{($^{\circ}$ E from N)}}
\startdata
NGC~4656UV & Ellipse & 12:44:15.68 & +32:17:00.4 & 180.3x115.8 & 36.36\\
NGC~4656 & Ellipse & 12:43:53.61 & +32:08:46.0 & 390.0x142.2 & 40.00\\
NGC~4656-Hook & Box & 12:44:08.82 & +32:12:36.8 & 127.2x82.8 & 100.0\\
NGC~4656-Center & Box & 12:43:57.30 & +32:09:47.6 & 376.2x162.5 & 30.00\\
NGC~4656-Tail & Box & 12:43:40.62 & +32:05:32.9 & 328.5x202.8 & 50.00\\
\enddata
\tablenotetext{a}{For ellipses: the semi-major and semi-minor axes. For boxes: the length of the sides.}
\label{tab:regions}
\end{deluxetable*}

\subsubsection{Surface gradients}
When computing photometry for sources near the detection limit, as necessitated here for most filters, small background fluctuations can add significant systematic errors to the results. These background gradients are independent of the random errors used in our S/N thresholds and act to increase the true error in wavelength bands where an object has a small number of counts relative to the background level. 

To combat this problem we fit and subtract polynomial surfaces to all images before they are photometered. Generally when subtracting a background fit, we seek a solution that converges at large polynomial order. However, the commonly used scheme of masking sources tends to fail at high orders: restricting the location of points to perform the fit leads to poor surface subtraction in areas where a significant number of the pixels are masked, and increases errors overall when masks force non-optimal point selection (especially for polynomial surface fits).

\begin{figure}
\begin{centering}
\includegraphics[scale=0.5]{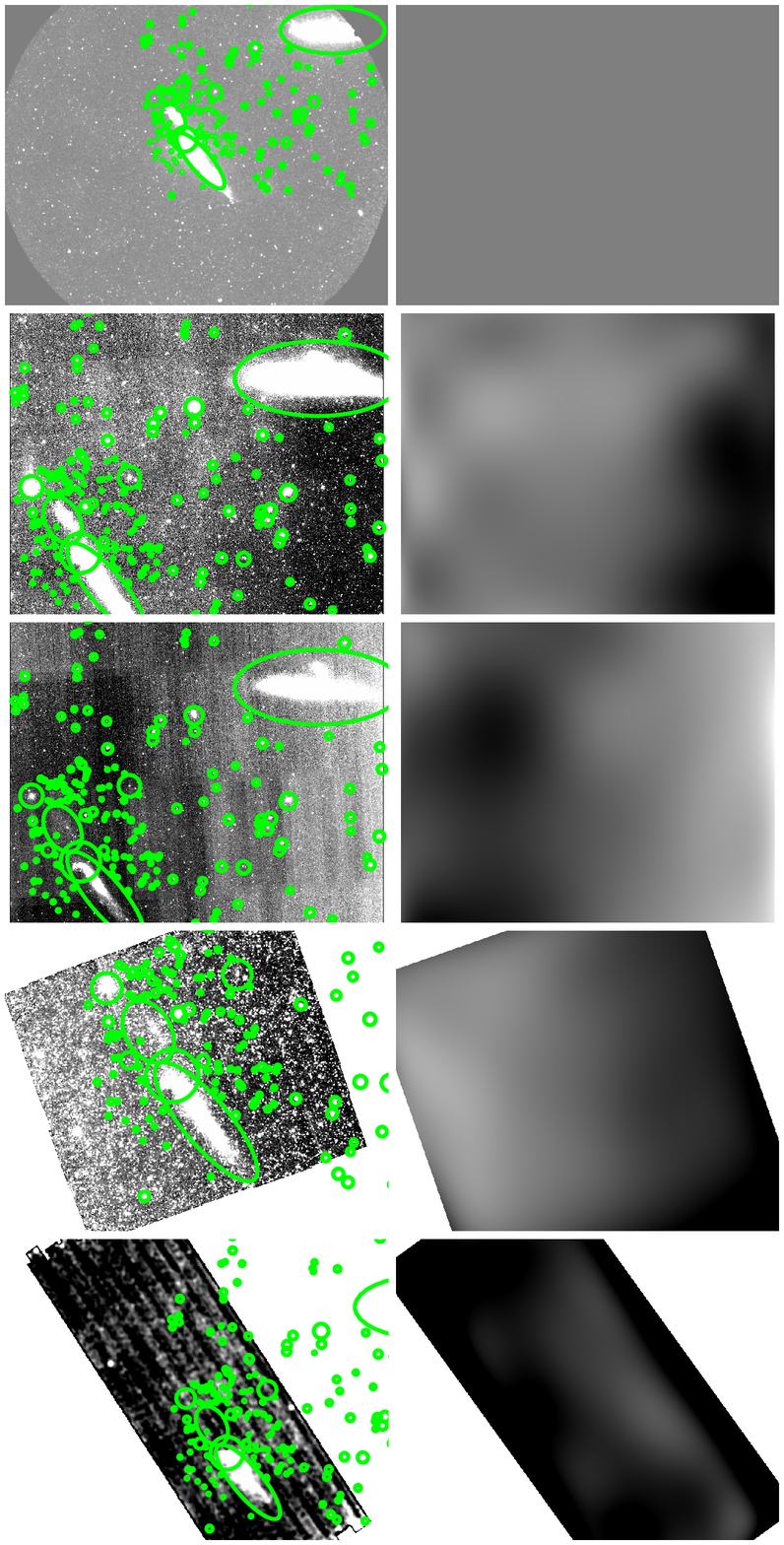}
\caption{Demonstration of our surface-subtraction technique. Left column: original image. Right column: 9th order surface fit using our technique. From top to bottom the rows are: GALEX FUV, SDSS $r$, SDSS $z$, IRAC 3.6 $\mu$m, and MIPS 70 $\mu$m. Green ellipses are the masking apertures used for the surface fits. All images have been rotated to North-up, East-left orientation and smoothed by a gaussian with $\sigma$=6 pixels to highlight faint structure. Our algorithm accurately fits the surface gradients for all of the images at high polynomial order, even for images with significant padding (e.g. the FUV, 3.6 $\mu$m, and 70 $\mu$m images) and/or large mask regions (e.g. the $r$ and $z$ images).}
\label{fig:skysubcompare}
\end{centering}
\end{figure}

To solve these problems we took an iterative approach to build surface fits that were both accurate and stable at high polynomial orders. We started by fitting a zeroth order Lagrange polynomial (a constant offset) to the unmasked data using the IRAF task {\it imsurfit}. This fit is minimally sensitive to extended sources on the image.  We replaced the pixel values of all masked regions by the value of the fitted zeroth order background and the resulting image is run through imsurfit with order = 1. This process is repeated while increasing the polynomial order until the change in the surface fit between orders asymptotes to a minimum value, which we find happens after 8 orders for most images (although for the 2MASS H-band we go to 19th order due to the larger variations in the sky background in that bandpass). 

For the GALEX, SDSS, and 2MASS images, the change between polynomial orders asymptotes at $\le$0.05\%. Because the IRAC and MIPS mosaics are not perfectly rectangular our algorithm's performance is not as good, with variations $\le$1.5\%, although we note that the largest variations are confined to the corners of these images and that the surface fit generally has less variation in the central parts of the mosaic. As Figure \ref{fig:skysubcompare} shows, our method is robust regardless of the fitting order.

While our method is a large improvement over non-iterative fits, there are still limitations to surface-removal. When the surface gradients are variable over scales comparable to the objects to be photometered, it is impossible to accurately subtract them without also subtracting the object. Fortunately in most of our images the gradients are on large scales and/or both NGC~4656UV and all parts of NGC~4656 have significantly higher counts than the background fluctuations. However, there are two exceptions to this which will be discussed further in \S \ref{SF} and \S\ref{GALEV}.

\section{Results}
\label{results}

We present our results from archival and survey observations in two parts.
First, we present the VLA 21 cm observations and evidence from the cold
neutral \hi\ gas that NGC~4656/NGC~4656UV is a disturbed system.  Then we
present the UV/optical/infrared imaging and modeling of the spectral energy
distribution (SED) to demonstrate that the age and metallicity of the NGC~4656UV stellar system is consistent with the interaction history of the group, providing a self consistent picture for the creation of the tidal dwarf galaxy candidate.

\subsection{Evidence for the interaction history of NGC~4656}

The distribution and kinematics of \hi\ gas in the environment of the NGC~4656/NGC~4631 group provides the strongest evidence that the two edge-on spiral galaxies have undergone a recent interaction.  On the galaxy group scale, \hi\ ``fingers'' extend from NGC~4631 towards NGC~4656, as if the \hi\ gas has been stripped out of the former as the result of a gravitational encounter between the two galaxies.  This has been well documented by two sets of authors, and we refer the reader to Figures 1 and 3 of \citet{Weliachew78} and Figures 4 and 7 of  \citet{Rand94} for a perspective of \hi\ across the entire group.

\begin{figure*}
\begin{centering}
\includegraphics[scale=0.6]{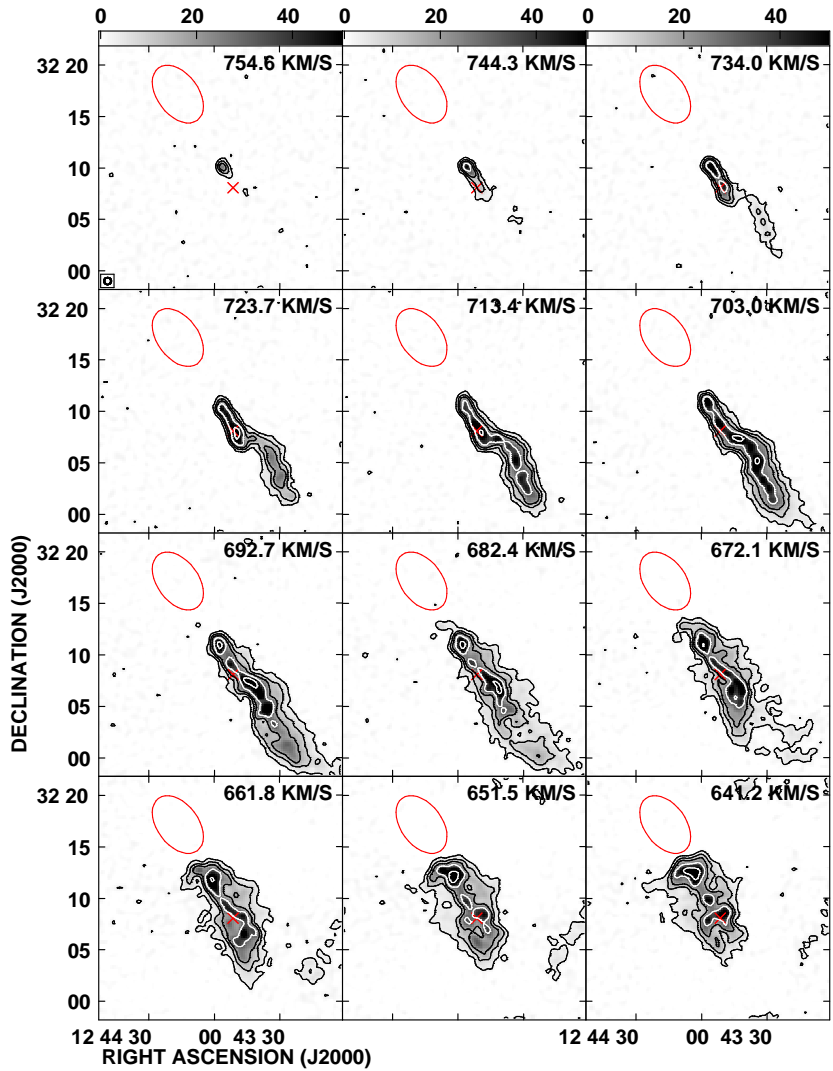}
\caption{Channel maps of NGC~4656 covering 754.6 km s$^{-1}$ to 641.2 km s$^{-1}$ heliocentric radio velocity.  Data are plotted at full spatial and spectral resolution.  The beam size is drawn in the lower left corner of the first panel. Contours are plotted for 3, 10, 20, 40, and 80$\sigma$ where $\sigma=5.03\times10^{18}$ cm$^{-2}$.  Grayscale is in units of mJy beam$^{-1}$. The ``X'' marks the center of the galaxy found through our modeling of the rotation curve.  The red ellipse marks the position of the TDG candidate. Note the significant amount of gas to the northeast of the center in all panels, which is counterrotating with respect to the rest of the galaxy.} 
\label{chans1}
\end{centering}
\end{figure*}

\begin{figure*}
\begin{centering}
\includegraphics[scale=0.6]{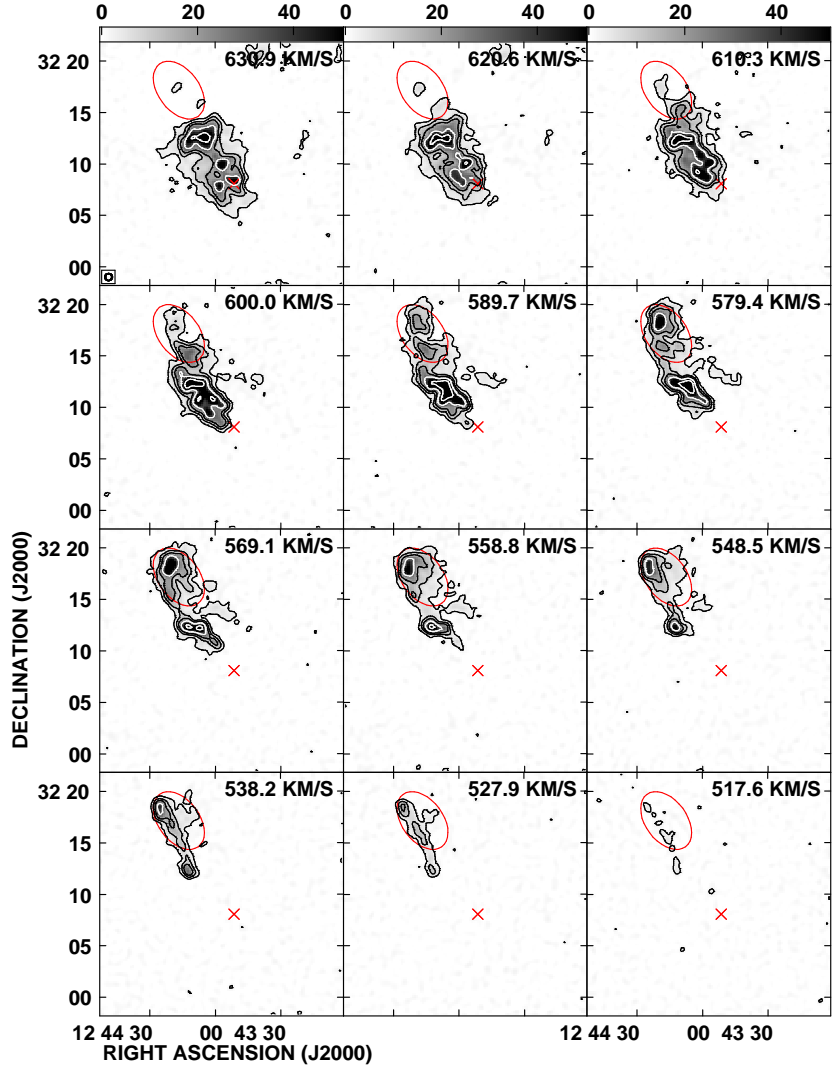}
\caption{As in Figure \ref{chans1} but covering 630.9 km s$^{-1}$ to 517.6 km s$^{-1}$ heliocentric radio velocity.  \hi\ associated with the TDG is found most strongly in channels covering 600 km s$^{-1}$ to 527.9 km s$^{-1}$.  Between this and Figure \ref{chans1}, it is clear the galaxy is rotating, but is very asymmetric.} 
\label{chans2}
\end{centering}
\end{figure*}

\begin{figure}
\begin{centering}
\plotone{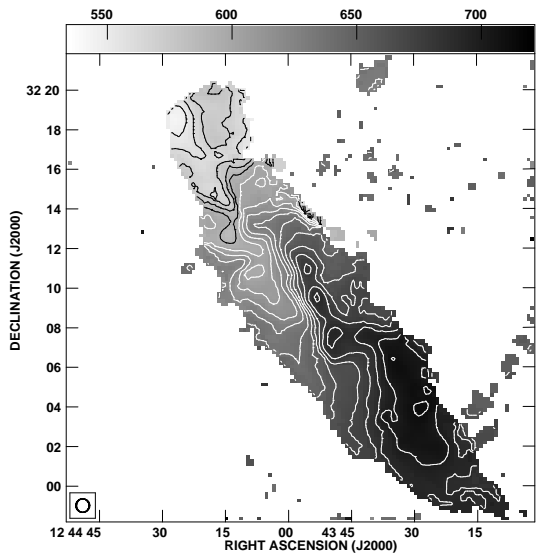}
\caption{The intensity weighted velocity (moment 1) map of NGC~4656 and TDG candidate, NGC~4656UV.  Emission has been clipped at 7$\sigma$, and contours range over $540-720$\kms\ in steps of 10\kms.  We note there is a continuous velocity gradient from the parent galaxy across the TDG companion.  The twisted isovelocity contours show that the parent galaxy is clearly disturbed. The extraplanar emission to the northwest and southwest may be real emission associated with the \hi\ ``fingers'' between NGC~4656 and NGC~4631 reported by \citet{Weliachew78} and \citet{Rand94}.}
\label{mom1}
\end{centering}
\end{figure}

We use the archival VLA observations to investigate the kinematics and gas morphology of NGC~4656.  We present the \hi\ data cube in the form of velocity channel maps in Figures \ref{chans1} and \ref{chans2}, total intensity map (moment 0) contours overlaid on a composite image in Figure \ref{imagewcontours}, and an intensity weighted velocity (momen 1) map in Figure \ref{mom1}). The moments are calculated independently for each pixel. The intensity weighted velocity map was created by clipping the emission below 7$\sigma$ of our RMS sensitivity in order to avoid large outliers. The systemic velocity of NGC~4656 is $645\pm3$ km s$^{-1}$, while NGC~4656UV has a systemic velocity of about 570\kms.

The channel maps show the presence of warping in the disk, as well as counterrotating and extraplanar gas.  In each panel the ``X'' marks the center of the galaxy found through modeling of the rotation curve.  In channels 754.6\kms\ through at least 723.6\kms\, we see gas northeast of the center of the galaxy which is counterrotating with respect to the disk.  Channels 723.7\kms\ through 703.0\kms\ show warping in the outer portion of the disk.  Channels 600.0\kms\ through 548.5\kms\ show evidence of extraplanar gas (north of the ``X''; identified as a ``worm'' by \citealt{Rand94}).  The total \hi\ intensity map (Figure \ref{imagewcontours}) shows the neutral hydrogen gas extends beyond the optical plane of the galaxy both radially and vertically.   In \hi\, the TDG (red ellipse in Figures \ref{chans1} and \ref{chans2}) appears as a continuous extension of the galaxy to the northeast.  We used our knowledge of the optical morphology to exclude the gas in this region from our modeling of the rotation curve of the parent galaxy.

\subsection{Modeling the rotation curve of NGC~4656}

\begin{figure}
\begin{centering}
\includegraphics[scale=0.4]{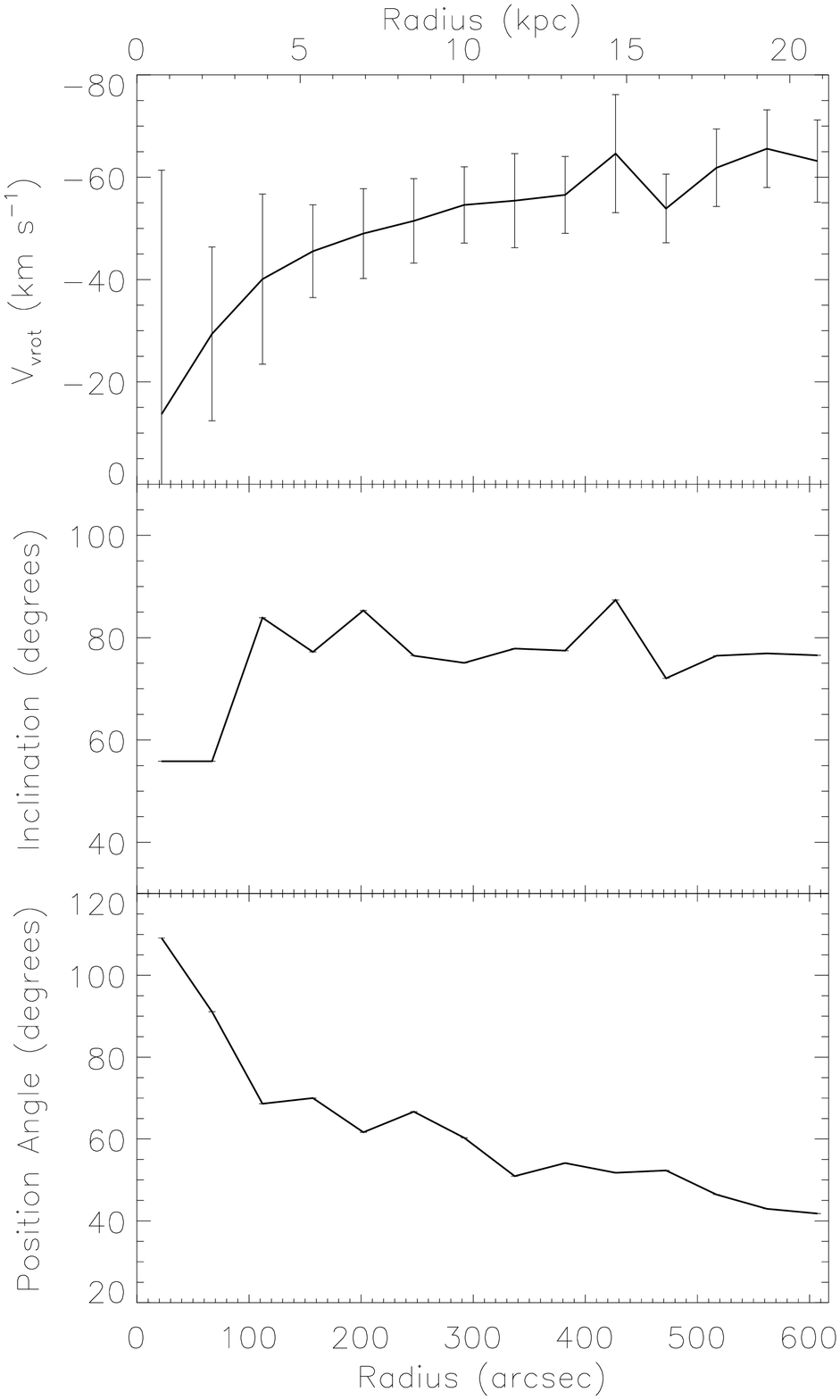}
\caption{Top panel: rotation curve for NGC~4656 derived from a tilted ring model.  Middle panel: varying inclination angle with radius.  Bottom panel: varying position angle with radius.  The rotation curve is shallower than suggested by the velocity channel maps and does not capture the presence of anomalous or counterrotating gas.  The position angle declines by about 60 degrees from the center to the outskirts of the galaxy.  This change in position angle is the model's attempt at fitting the twisted isovelocity contours apparent in the moment 1 intensity weighted velocity map.  While the galaxy is disturbed at the center, the shape of these isovelocity contours is likely due to the counterrotating gas instead of a warp in the disk.} 
\label{rotcur}
\end{centering}
\end{figure}

We attempted to fit a rotation curve to NGC~4656 using a tilted ring model following the procedure of \citet{Begeman89} and \citet{Hess09} with mixed success.  The top panel of Figure \ref{rotcur} shows the results of fitting a rotation curve to concentric rings assuming an ideal, flat disk with constant inclination and position angle.  We derived a systematic velocity of $645\pm3$\kms\ and inclination of approximately $79^{\circ}\pm3^{\circ}$.  We see that the rotational velocity rises slowly with radius, reaching a maximum of $\sim65$\kms\ before roughly flattening out.  From the rotation curve, we derived a total dynamical mass of $1.9\times10^{10}$ M$_{\odot}$  for NGC~4656.

The middle and bottom panels of Figure \ref{rotcur} demonstrate the quality of our assumption that the galaxy is an ideal disk by plotting how the inclination and position angle, respectively, vary with radius when we fix the rotation velocity of each ring to that found in the simple, ideal case. 
The ideal case fails to match all the features observed in the data: it cannot account for the twisted isovelocity contours in the intensity weighted velocity map (Figure \ref{mom1}) at the center of the main galaxy, indicating that NGC~4656 deviates from regular rotation.  Further modeling shows the twisted isovelocity contours can be fit with a position angle which varies with radius, as demonstrated in the bottom panel of Figure \ref{rotcur}.   However, when we examine the data cube, the center of the galaxy is dominated by non-circular motions and clumps of counterrotating \hi\ gas at ``forbidden'' velocities, seen in velocity channel maps, which twist the isovelocity contours.

\begin{figure*}
\begin{centering}
\plotone{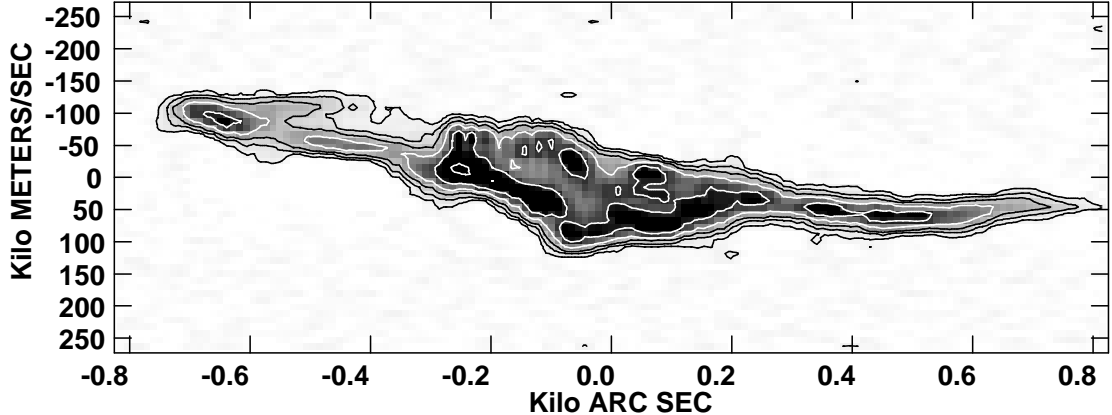}
\caption{A position-velocity slice taken from the \hi\ data cube along the major axis of the parent galaxy.  The TDG candidate NGC~4656UV is the density enhancement to the left at approximately -0.65 kiloarcseconds. The parent galaxy appears to have some rotation, however the \hi\ gas is clearly disturbed.  We believe this is in part due to a previous gravitational interaction and to the vigorous star formation taking place in the parent galaxy.}
\label{major}
\end{centering}
\end{figure*}

Figure \ref{major} shows a position-velocity (p-v) slice through the \hi\ data cube along the major axis of the galaxy.  
This slice is atypical compared to what is usually observed in spiral galaxies, which exhibit a quickly rising, then flat rotation curve (e.g. \citealt{Sofue01} and references therein).
In the inner 300 arcseconds ($\sim10$ kpc) the gas exhibits a huge velocity dispersion ($\sim100-150$\kms) compared to the overall change in the rotation velocity with radius (of order $60$\kms).  The \hi\ gas is not concentrated at the systemic velocity of the galaxy, but at higher and lower velocities as if the gas is expanding in the radial direction.  Given the shallow potential well inferred from the slowly rising rotation curve and that NGC~4656 is a fairly low mass spiral (100 times less massive than the Milky Way), the modest star formation rate (a factor of 4-5 times less than the Milky Way, see \S\ref{sec:sfr}) may be sufficient to blow gas out of the center of the galaxy.

We believe the disturbed \hi\ gas and gas at large heights above the disk of NGC~4656 are due to a combination of the galaxy-galaxy interaction and the current episode of vigorous star formation. Extraplanar and counterrotating gas has been observed in several edge-on and inclined galaxies (e.g. NGC~891, \citealt{Oosterloo07}; NGC~2403, \citealt{Fraternali02}; NGC~2997, \citealt{Hess09}).  The presence of this gas may be indicative of active star formation, however the largest \hi\ clouds are likely evidence of accretion onto the galaxy---either of primordial origin (e.g. \citealt{Hess09}; \citealt{Wilcots98}), or as the result of tidal material falling back onto the disk, such as has been modeled by \citet{Hibbard95}.

\subsection{HI gas in NGC~4656UV}
\label{higas}

Figure \ref{imagewcontours} shows an \hi\ envelope spatially coincident with NGC~4656UV, although offset from the apparent UV-optical center of the TDG candidate.  This \hi\ envelope is connected to its parent galaxy by a smooth velocity gradient in the \hi\ gas: the p-v slice intersects the TDG candidate, identified as the density enhancement around $-100$\kms\ (Figure \ref{major}). 

We interpret the \hi\ density enhancement and the velocity gradient as evidence of the \hi\ gas having been pulled out of NGC~4656 during an interaction.  The \hi\ envelope serves as the gas reservoir out of which NGC~4656UV is currently forming stars.  Summing the \hi\ emission in this envelope yields a total flux of $2.7\pm0.5$ Jy.  At 7.2 Mpc this corresponds to a total \hi\ mass of $3.8\times10^{8}$ M$_{\odot}$ for the TDG candidate, consistent with the range of \hi\ masses in the \citet{Duc99} sample.  
The peak \hi\ column density towards the TDG is $1.3\times10^{21}$ cm$^{-2}$, however, the column density coincident with star formation across the rest of TDG is below the supposed threshold of 10 M$_{\odot}$ pc$^{-2}$ for the disk instability criterion \citep{Martin01}.  At the lowest level, we observe star formation occurring at gas surface densities of 1.6 M$_{\odot}$ pc$^{-2}$.  It is important to note that $\Sigma_{gas}$ here is based only on \hi\ observations, neglecting the molecular gas component which would raise the total gas surface density.  Nonetheless, the TDG candidate contributes to a growing body of evidence that star formation can occur in low density environments.

The \hi\ density enhancement we associate with NGC~4656UV is evident in previous observations, however \citet{Rand94} proposed it was one side of an \hi\ ring encircling the optically bright edge on galaxy.  We consider this and other possible explanations in \S\ref{discussion}. The identification of the UV-bright/optically faint stellar component to the northeast of the galaxy (NGC~4656UV) is motivation for reconsidering the gaseous structure.  There is no stellar counterpart to the southwest portion of the proposed \hi\ ring.  

\subsection{Spectral energy distribution of NGC~4656UV: recent star formation and dust limits}
\label{SF}

\begin{deluxetable*}{c c c c c c}
\tabletypesize{\footnotesize}
\tablewidth{0pt}
\tablecaption{Integrated Photometry (mJy)}
\tablehead{\colhead{Band} & \colhead{NGC 4656} & \colhead{Hook} & \colhead{Center} & \colhead{Tail} & \colhead{NGC 4656UV}}
\startdata
FUV&$77.4 \pm 3.8 $&$24.9 \pm 1.2$&$45.4 \pm 2.2 $&$9.89 \pm 0.48 $&$3.14 \pm 0.15$\\
NUV&$77.1 \pm 2.1 $&$22.6 \pm 0.60$&$ 48.8 \pm 1.3 $&$ 9.23 \pm 0.25 $&$3.22 \pm 0.086$\\
u&$91.0 \pm 0.12 $&$22.3 \pm 0.035$&$ 63.2 \pm 0.074 $&$ 10.1 \pm 0.064 $&$2.80 \pm 0.061$\\
g&$165 \pm 0.072 $&$32.1 \pm 0.029$&$ 122 \pm 0.057 $&$ 17.5 \pm 0.030 $&$5.67 \pm 0.025$\\
r&$178 \pm 0.094 $&$32.1 \pm 0.032$&$ 137 \pm 0.069 $&$ 17.3 \pm 0.044  $&$5.56 \pm 0.041$\\
i&$172 \pm 0.14 $&$28.2 \pm 0.040$&$ 139 \pm 0.093 $&$ 14.9 \pm 0.067 $&$5.29 \pm 0.071$\\
z&$153 \pm 0.46 $&$27.4 \pm 0.12$&$ 134 \pm 0.28 $&$ 6.81 \pm 0.28 $&$10.0 \pm 0.26 \pm 37^{b}$\\
J&$199 \pm 8.2 $&$32.1 \pm 2.0$&$ 159 \pm 4.9 $&$ < 50.5 $&$<58.8$\\
H&$206 \pm 14 $&$< 34.6$&$ 178 \pm 8.3 $&$ < 85.6 $&$<113$\\
K&$159 \pm 13 $&$< 32.4$&$ 126 \pm 7.8 $&$ < 80.0 $&$<87.2$\\
3.6&$80.2 \pm 0.027 $&$11.4 \pm 0.0079$&$ 64.1 \pm 0.019 $&$ 6.38 \pm 0.014$&$2.39 \pm 0.018$\\
4.5&$58.1 \pm 0.035 $&$8.65 \pm 0.0097$&$ 46.4 \pm 0.023 $&$ 4.10 \pm 0.019 $&---$^{a}$\\
5.8&$63.0 \pm 0.17 $&$9.22 \pm 0.050$&$ 48.5 \pm 0.11 $&$ 10.0 \pm 0.10 $&$<0.941$\\
8.0&$99.4 \pm .17 $&$17.7 \pm 0.043$&$ 70.1 \pm 0.10 $&$ 6.61 \pm 0.097 $&---$^{a}$\\
24&$530 \pm .31 $&$135 \pm 0.088$&$ 387 \pm 0.19 $&$ 17.4 \pm 0.18 $&$9.51 \pm 0.17 \pm 23^{b}$\\
70&$9850 \pm 100 $&$2290 \pm 48$ &$ 6760 \pm 82 $&$ 493 \pm 23 $&$<122$\\
160&$11800 \pm 170$& $1890 \pm 65$ &$ 841 \pm 140$ &$ 669 \pm 47$&$<388$
\enddata
\tablenotetext{a}{Significant portions of NGC~4656 were off the IRAC 4.5 and 8.0 $\mu$m tiles. To avoid inaccuracies from making assumptions about the morphology of the un-observed regions (as well as issues with surface gradients along the edges of the images) we do not photometer NGC~4656UV in these bands.}
\tablenotetext{b}{Significant small-scale surface gradients are co-spatial with the position of NGC~4656UV in the $z$ and 24 $\mu$m images, which caused the random errors to be much smaller than the true errors. For these two detections we first report the random error, then an estimate of the systematic error based on the amplitude of other small-scale surface gradients in the images.}
\label{tab:photometry}
\end{deluxetable*}

A summary of the UV, optical, and infrared photometry for NGC~4656UV, NGC~4656, and various morphologically distinct star forming regions within NGC~4656 is shown in Table \ref{tab:photometry}.  An upper limit indicates the flux that would be present from a source with a 10$\sigma$ detection. The uncertainties are based on random errors only and do not include systematic effects (e.g. un-removed surface gradients). Therefore our uncertainties should be treated as lower bounds on the true uncertainty. 

The $z$-band and 24 $\mu$m images appear to be significantly affected by surface gradients. Photometry of the 24 $\mu$m image resulted in a better than $10\sigma$ detection for NGC~4656UV; however a visual inspection showed a structure more consistent with small-scale background variations seen elsewhere in the image than with the morphology of NGC~4656UV observed in other bands, so we do not use this ``detection'' in our analyses. There is clearly $z$ band emission from NGC~4656UV in the SDSS image, but it appears to be super-imposed over a surface gradient with amplitude comparable to NGC~4656UV. We discuss this detection more thoroughly in \S\ref{sed}. For these two bands we also estimated the magnitude of the surface gradients on the image and report these as additional errors on the integrated flux of NGC~4656UV in Table \ref{tab:photometry}.

\subsubsection{Spectral energy distributions}
\label{sed}
\begin{figure*}
\begin{centering}
\plottwo{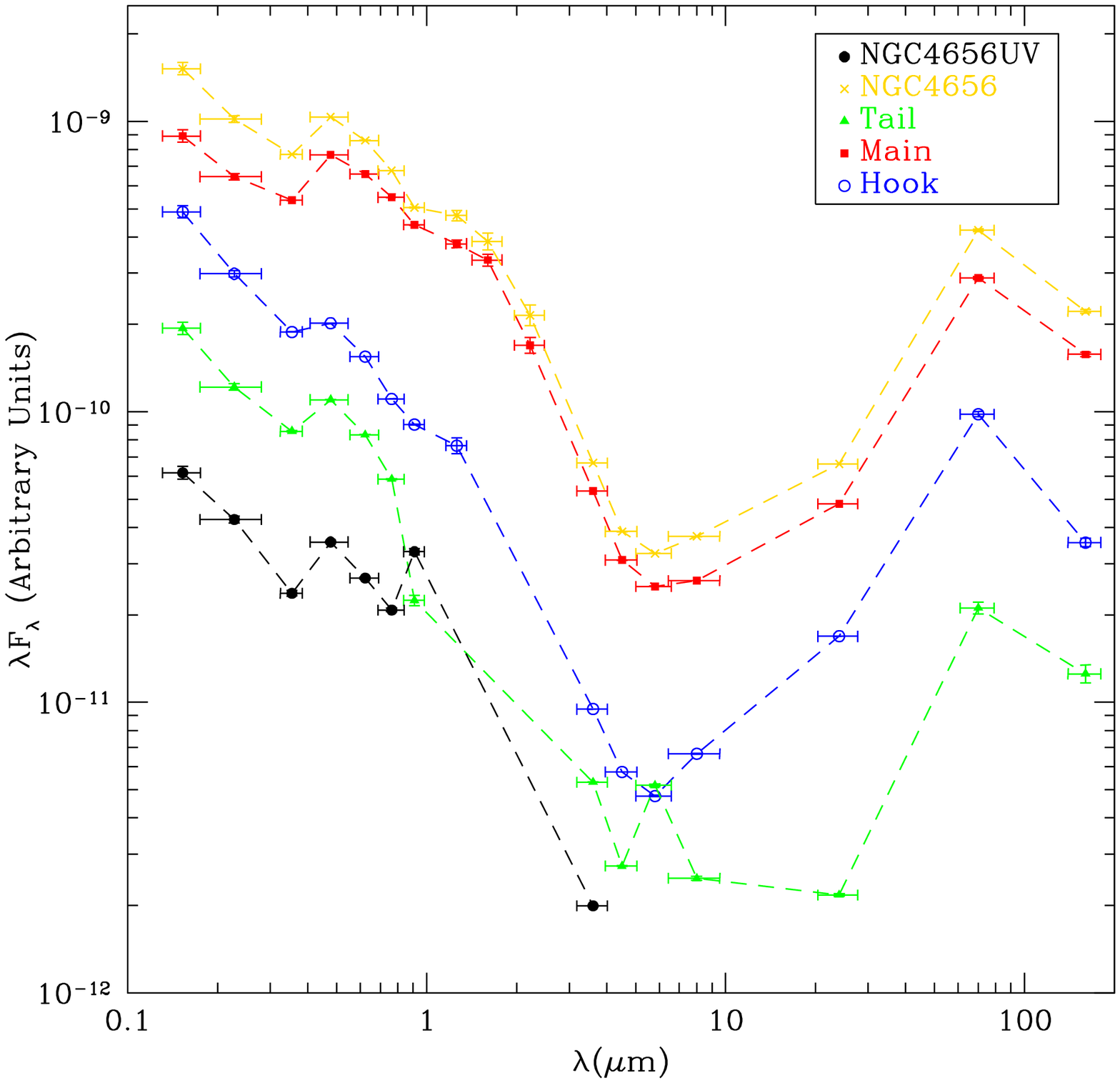}{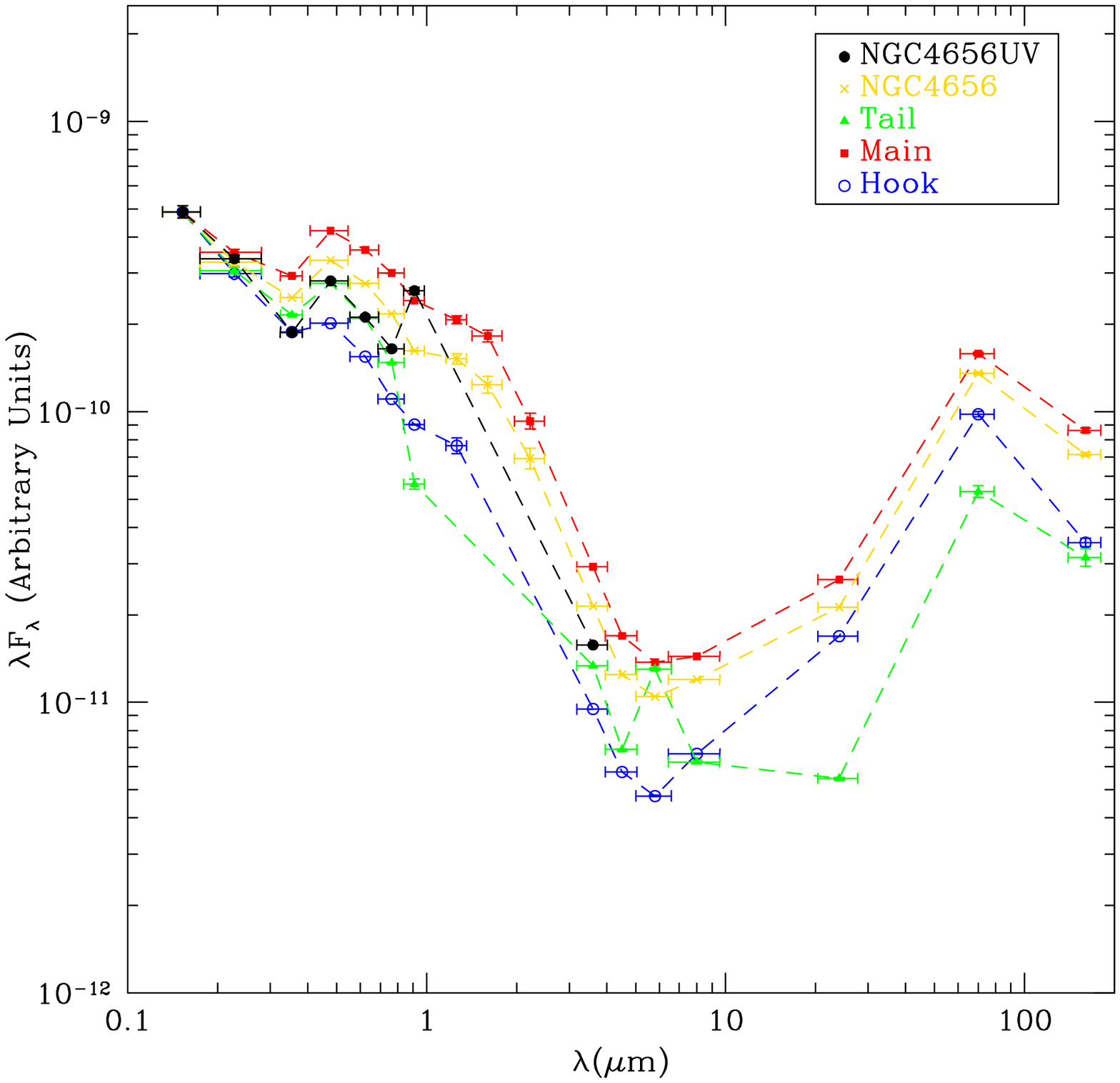}
\caption{SEDs of all regions of NGC~4656 and NGC~4656UV. The left panel preserves the relative fluxes while the right panel scales all the SEDs to have identical FUV fluxes. Horizontal error bars denote the width of the filters. The entire NGC~4656-4656UV system is clearly undergoing a similar-age burst of star-formation.}
\label{fig:seds}
\end{centering}
\end{figure*}

We plot the SED of NGC~4656UV with the SEDs of its parent galaxy in the left panel of
Figure \ref{fig:seds}. To facilitate comparisons, we also
show the SEDs scaled to have the same FUV flux in the right panel of Figure
\ref{fig:seds}. All regions of NGC~4656 are actively forming stars, and have
similar UV-optical colors to NGC~4656UV (with the exception of the $z$-band,
discussed below). The SED of NGC~4656UV is very similar to that of NGC~4656's
tail, the other low-density star formation environment in this system. Both
these systems have similar ratios of FUV-NUV-$u$ band fluxes, and steep optical
slopes. The hook also exhibits this morphology, although with a nearly
non-existent Balmer break, in contrast to the strong breaks of NGC~4656UV and
the tail of NGC~4656. The main region (and, due to the main region's significantly higher
brightness than the tail or the hook, NGC~4656 as a whole) has a shallower
slope in both the UV and the optical, indicative of old stellar populations present in the center of the galaxy. All parts of NGC~4656 contain
cold dust, although the tail appears to be deficient in 70 $\mu$m emission compared to the rest of the galaxy.

The $z$-band detection in NGC~4656UV is odd, considering how similar the rest of its UV-optical SED is to the postulated parent galaxy. Significant excess flux in the $z$-band would be a strong indicator of an older stellar population, one either formed in-situ (if NGC~4656UV is a pre-existing dwarf galaxy) or ripped from NGC~4656 at the same time as the neutral gas currently powering the ongoing episode of star formation (if NGC~4656UV is a TDG). A visual inspection of the $z$-band image clearly shows a signal that is consistent spatially with NGC~4656UV, however the detection is at small DN/pixel and there are similar-magnitude surface gradients nearby, similar to the 24 $\mu$m band (see row 3 of Figure \ref{fig:skysubcompare}, and note especially the vertical bands on the image).  We believe that these gradients are responsible for artificially boosting the $z$-band flux given the otherwise strong similarity between the SEDs of NGC~4656UV and NGC~4656, but in \S\ref{GALEV} we consider the consequences of it being a real feature in NGC~4656UV's SED. 

\subsubsection{Star formation rates}
\label{sec:sfr}
We compute NGC~4656UV's star formation rate (SFR) using the conversion from FUV flux given by \cite{Kennicutt98}:
\begin{equation}
SFR (M_{\odot}\ yr^{-1})=1.4\times10^{-28} L_{\nu} (erg\ s^{-1}Hz^{-1}).
\label{eq:sfr}
\end{equation}
At the assumed distance of 7.2 Mpc, NGC~4656UV's FUV luminosity is $1.95\pm 0.09\times10^{26}$ erg s$^{-1}Hz^{-1}$.  Therefore, we calculate a star formation rate of $0.027 \pm 0.001$ M$_{\odot}$ yr$^{-1}$ for the TDG candidate. This is consistent with the mean star formation rate of TDGs according to \cite{Duc99}. For the parent galaxy, NGC~4656, we compute a SFR of $0.666 \pm 0.03$ M$_{\odot}$ yr$^{-1}$, roughly 25 times greater than that of NGC~4656UV.  The FUV SFR for NGC~4656 is comparable to literature estimates of the SFR from H$\alpha$, but roughly a factor of two smaller than the far-infrared SFR and the 1.4 GHz radio SFR \citep{Mapelli10}. This discrepancy may be caused by dust attenuation on the UV emission of NGC~4656, a source of uncertainty noted by \cite{Kennicutt98}; however, we will show in \S\ref{GALEV} that evolutionary synthesis models predict very low amounts of dust in NGC~4656UV, implying that attenuation corrections on the SFR of the TDG candidate should be small.

The \cite{Kennicutt98} conversion formula has been optimized for relatively stable, long-lived, quiescent galaxies. Additionally, the calibration of Equation (\ref{eq:sfr}) depends on the initial mass function and stellar library chosen for modeling this relationship. As NGC~4656UV is a very different star-forming environment than the galaxies used to obtain Equation (\ref{eq:sfr}) the SFR computed using this formula should be treated with caution.

\subsubsection{Should we expect to see dust in NGC~4656UV?}
One of the most obvious differences between the SEDs of NGC~4656 and NGC~4656UV is the lack of dust detected from mid- and far-IR emission in NGC~4656UV.
However, it is unclear whether the non-detections are due to a true lack of dust or the faintness of the TDG candidate. Therefore, we attempt to estimate what we should expect to observe for NGC~4656UV based on its UV and optical properties.

While all portions of NGC~4656 have similar SEDs in the UV and optical to NGC~4656UV, NGC~4656's tail is the most alike. If we assume that this similarity extends into the mid-IR, then using the same ratio of NUV to 70 $\mu$m flux as the tail of NGC~4656, we predict that NGC~4656UV should have a flux of 0.172 Jy at 70 $\mu$m: roughly 1.5 times the 10$\sigma$ detection limit. We use the NUV flux in this ratio because both bands should trace the ongoing star formation event, however we note that no ratio of wavelengths predict a flux below our 10$\sigma$ detection limit.  The estimated NUV/160 $\mu$m flux ratios are, however, consistent with a non-detection for NGC~4656UV. These ratios indicate that NGC~4656UV has a lower cold dust density than even the most diffuse regions on NGC~4656. 

Given the very active star formation in NGC~4656UV, it is also valuable to estimate the expected flux from polycyclic aromatic hydrocarbons (PAHs). In a sample of star forming regions formed in collisional debris, \citet{Boquien09} report 8.0 $\mu$m detections in all systems, indicating that many tidal features exhibit emission from small grains of PAHs (although Holmberg IX, a nearby TDG candidate that will be discussed in more depth in \S\ref{comparetdgs}, was not detected in 8.0 $\mu$m by \citealt{Dale07}). PAH emission has also been detected in TDGs at several wavelengths by \citet{Higdon06}. Unfortunately, since the archival IRAC 8.0 $\mu$m image only contains about half of NGC~4656UV, a direct comparison of 8.0 $\mu$m fluxes with UV emission using the formulae given by \citet{Boquien09} is impossible. However, the general lack of dust emission in the IRAC bands suggests an overall under abundance of PAHs. This may be due to NGC~4656UV's very low metallicity, consistent with our SED analysis and GALEV modeling (detailed in \S\ref{GALEV}).

\subsection{Evolutionary synthesis modeling}
\label{GALEV}
In order to gain more insight into NGC~4656UV's star-formation history, age, metallicity, and extinction we created a set of stellar population synthesis models using GALEV \citep{Kotulla09}. Being chiefly interested in the ongoing starburst we chose an exponentially declining burst star formation rate with an e-folding time of 200 Myr. We considered a set of five metallicities ranging from [Fe/H] = $-1.7$ to 0.0 (solar), including a model with a ``chemically consistent metallicity''. The ``chemically consistent'' model starts with the burst at primordial metal abundance, then tracks the evolving gas-phase metal abundance as the stellar populations age, die, and return metals into the ISM (see \citealt{Kotulla09} for more information). We include this model as the most limiting case of low metal enrichment.  We varied the amount of dust, using a Calzetti attenuation model with extinction up to E(B-V) = 0.5 mag \citep{Calzetti00}. 

\begin{figure*}
\begin{centering}
\includegraphics[scale=0.6]{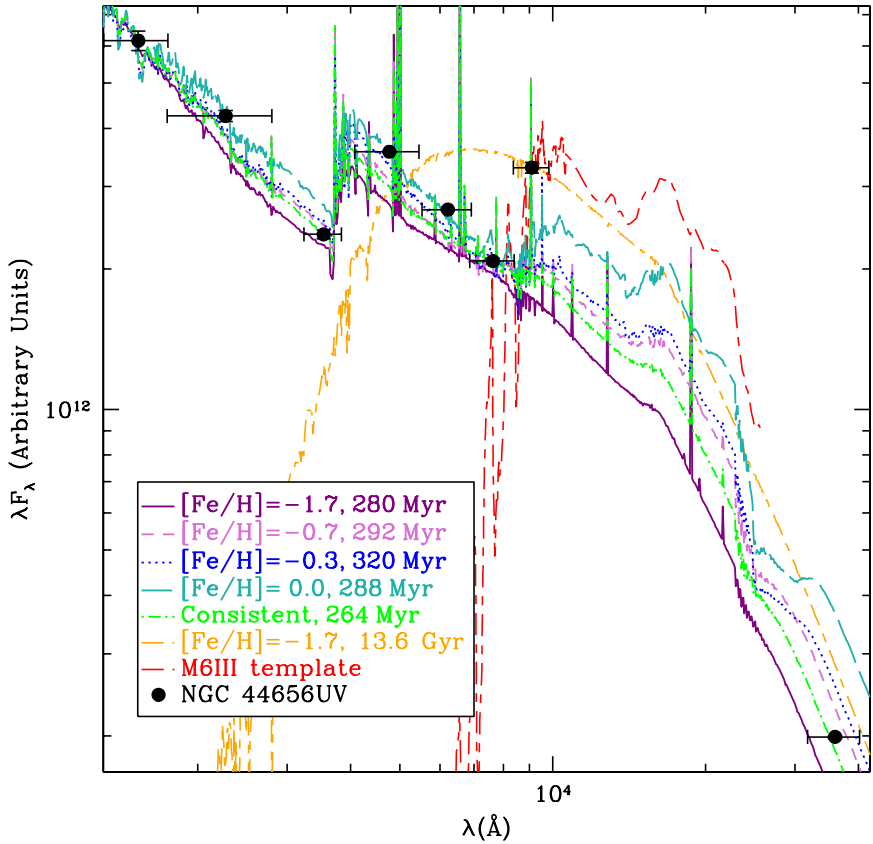}
\caption{GALEV fits to the bluest six bands of NGC~4656UV's SED. Data from NGC~4656UV are shown as black points, while the best fitting GALEV models to the UV+ugri data for each metallicity bin are shown as colored lines. Blue, green, and purple lines denote the best fitting models, while the orange line is for the most metal-poor model at an age of 13.6 Gyr. The most metal poor model fits to the UV and optical data well reproduce both the UV and optical data but are also consistent with the 3.6 $\mu$m detection. However, attempting to fit the $z$-band flux by adding a highly evolved stellar population fails, overpredicting the flux in neighboring bandpasses. The red line denotes the spectrum of an M giant template star from \citet{Bruzual93}, which would be the characteristic spectrum of any old stellar population that could fit the $z$-band flux without overpredicting $r$ and $i$ band emission.}
\label{fig:galevplot}
\end{centering}
\end{figure*}

For each model, fluxes are computed for all passbands, then a least-squares
fit is performed at each model age to fit the model fluxes to the data for
NGC~4656UV. To focus on the recent star formation we fit only to the UV+$ugri$ data, the
region of NGC~4656UV's SED which would be dominated by a young, recently
formed stellar population.  In Figure \ref{fig:galevplot} we plot the model
with the best fitting age for each metallicity, overlaid with the measured
flux of NGC~4656UV in each wavelength band.  Adding dust to the
models decreased the resulting age of the system.  For example, adding
E(B--V) = 0.1 mag of dust resulted in a younger best fitting age for the TDG by
a factor of 1.5---2, regardless of metallicity. However, the best fitting models
for all metallicities had E(B--V) $<$ 0.05 mag, with most models requiring zero
dust. As the metallicity increases there is a slight trend of increasing best
fit age: the best fit ages are 280 Myr at [Fe/H] = $-1.7$, 292 Myr at
[Fe/H] = $-0.7$, 320 Myr at [Fe/H] = $-0.3$, and 288 Myr at solar.
The chemically consistent model has the youngest age, 264 Myr. The best
fits overall are given by the chemically consistent and [Fe/H] = $-1.7$
models. At the best fit age the chemically consistent model has a
metallicity of [Fe/H] = $-2.36$, $\sim$0.65 dex lower than even the least abundant fixed
metallicity model. 

The apparent low metallicity of NGC~4656UV is surprising given that TDGs are traditionally metal-rich in relation to normal dwarfs. However, the system as a whole is metal-deficient with NGC~4656 having a metallicity of $-1.12$ dex \citep{Mapelli10}. NGC~4656UV's metallicity is $\sim$10 times smaller than that of its parent galaxy.  This may be explained if NGC~4656UV was formed out of gas stripped from the extreme outer disk of NGC~4656, which would be more metal-poor than gas closer to the center of the parent galaxy. While TDGs of all luminosities appear to have a fairly consistent Z$_{\odot}$/3 metallicity \citep{Duc99}, much higher than found for NGC~4656UV, the overall dearth of matter and metals in this group indicate that NGC~4656UV is forming in a different environment than most TDGs.

If we assume that the $z$-band excess is due to background gradients we find that while the models are minimized only to the bluest six detections, both the chemically consistent and [Fe/H] = $-1.7$ models do a fairly good job approximating the 3.6 $\mu$m flux. The chemically consistent model indicate that all of the 3.6 $\mu$m emission comes from starlight while the [Fe/H] = $-1.7$ model allows for a dust contribution of roughly 10-25\% of the total light. At [Fe/H] = $-0.7$ and higher the models over predict the 3.6 $\mu$m emission, a further indication that NGC~4656UV contains low metallicity gas. 

We investigate the possibility that the $z$-band excess is due to an older, more evolved population from a much earlier starburst by over plotting the SED of the [Fe/H] = -1.7 model for a 13.6 Gyr old burst on Figure \ref{fig:galevplot}. Even for this extremely old population, too much light is emitted in the $r$, $i$, and 3.6 $\mu$m bands to fit NGC~4656UV's SED. In order to fit the $z$-band signal without over predicting the emission in other bandpasses the old population would have to have the characteristic spectrum of a late M-type star (plotted in red in Figure \ref{fig:galevplot}). The inability of the 13.6 Gyr model to even approach a good fit of the data is another indicator that the $z$-band flux is aberrant. However, we note that even with the current data, we cannot conclusively rule out the possibility that there is an extremely old stellar population in NGC~4656UV. 

\section{Discussion}
\label{discussion}

While it is generally difficult to confirm that a galaxy is indeed a tidal dwarf without detailed stellar population studies (e.g. Holmberg IX, \citealt{Sabbi08}) or detailed dynamical modeling (e.g. \citealt{Hibbard95}), we favor NGC~4656UV's recent formation through a tidal interaction, most likely between NGC~4656 and NGC~4631. In the following sections we discuss the likelihood of other possible origin scenarios in the context of the kinematic and stellar data we have presented. Finally, we compare NGC~4656UV to observations of other star forming systems and tidal dwarf candidates.

\subsection{NGC~4656UV: the case for a tidal dwarf}

We postulate that NGC~4656UV formed out of gas stripped from the outskirts of NGC~4656 during a recent interaction.  The most likely culprit for this interaction is NGC~4631.  The difference in the line-of-sight velocities towards NGC~4656 and NGC~4631 is $\sim$40\kms, which is extremely low for a group velocity dispersion, and may suggest that most of the relative velocity between the two galaxies is in the plane of the sky.  With a projected separation of 36 kpc and assuming a typical group velocity dispersion of 200\kms, we estimate that the two spiral galaxies could have interacted as recently as 230 Myr ago, comparable to the post-starburst age estimates from the GALEV models. Detailed kinematic modeling is required to strengthen the case that NGC~4656UV was created by an interaction, but the \hi\ bridge observed by \citet{Weliachew78} and \citet{Rand94} and lend weight to NGC~4631 as the perturber.

Several features of the UV-IR SED and the neutral hydrogen maps imply a tidal origin for NGC~4656UV. The disassociated morphology of NGC~4656UV, seen both in \hi\ contours and especially in the UV stellar emission, indicates that NGC~4656UV is a separate entity from NGC~4656. NGC~4656's tail, while possessing a similar SED to NGC~4656UV, shows no signs of being disconnected from the main body of the galaxy. At the same time, the existence of a light bridge between NGC~4656UV and its putative parent as well as the continuity of the rotation of the neutral hydrogen gas between the systems implies a strong dynamical connection between them.  

If we assume the velocity gradient across the \hi\ density enhancement is due to rotation in the TDG candidate, we can use the measured baryonic mass, velocity, and size to estimate the virial ratio
\begin{equation}
\frac{T}{U}=\frac{RV^{2}}{3GM},
\end{equation}
where T and U are the kinetic and potential energies, respectively, and we have made the simplifying assumption that NGC~4656UV is a constant-density sphere. We compute a total \hi\ mass of $3.8\times 10^{8}$ M$_{\odot}$ for the TDG. Using the $g$-band magnitude, the estimated distance to the TDG, and a M/L ratio of 1, we estimate a stellar mass of $\sim 9\times 10^{7}$ M$_{\odot}$, giving a total \hi\ + stellar mass of $4.7\times 10^{8}$ M$_{\odot}$. Assuming a diameter of $296^{\prime\prime}$ ($\sim10.3$ kpc), equal to the average of the semi-major and semi-minor axes of our photometric aperture (large enough to contain most of the \hi\ emission), and a total velocity gradient of 73\kms, we estimate a virial ratio for NGC~4656UV of 1.13. This indicates that NGC~4656UV is right at the cutoff of being gravitationally unbound, perhaps unsurprising given that it appears to have been so recently liberated from NGC~4656 but has remained as a single entity. Combined with our estimate of the virial ratio, the large diameter of NGC~4656UV may indicate that the system is at least partially (if not fully) unbound. We note that this estimate is extremely rough: the density is not constant across the entire TDG, and the peak of the density distribution is not coincident with the center of the stellar distribution. This estimate also does not take into account kinetic energy due to a velocity dispersion, or expansion/contraction of the gas \citep{Bournaud04}. Additionally, even if some or all of NGC~4656UV is self-gravitating it is unclear whether or not this system will be fully ejected from NGC~4656 or will fall back onto its parent galaxy. 

We can also use these estimates of rotation velocity and size to estimate a total dynamical mass: 
\begin{equation}
M_{dyn}=\frac{RV^2}{G}.
\end{equation}
We estimate that $M_{dyn} \approx 1.6\times10^9$ M$_{\odot}$ for NGC~4656UV, only 3.4 times larger than the mass of the \hi\ gas + stars.  This discrepancy is similar to that found in other systems believed to be formed out of collisional debris (e.g. \citealt{Duc07,Bournard07}). Some of the missing mass may be accounted for by the presence of molecular gas (e.g. CO, H$_2$).  Regardless, it appears that NGC~4656UV has a relatively low ratio of dark-to-luminous matter compared to other dwarf galaxies (see \citealt{Ashman92} for a review), further supporting our hypothesis that NGC~4656UV formed from tidal interactions. 

Unfortunately, we cannot make a confident statement about the future of this star forming system and whether it might be long lived.  Although the data suggests that the system is not dark matter dominated, CO data would be even more powerful in revealing the internal kinematics of the system and determining whether it is gravitationally bound \citep{Braine01}.

\subsection{Comparison to other TDGs}
\label{comparetdgs}
It is clear from the photometry and \hi\ data that NGC~4656UV is vigorously turning gas into stars. We quantify this extreme star formation by comparing NGC~4656UV with a relevant sample from the literature. \cite{Smith10} selected 42 nearby pre-merger, interacting galaxy pairs and obtained broadband GALEX photometry for the galaxies as well as their tidal tails, bridges, and likely tidal dwarf candidates. Many of these features are very bright in the UV and nearly invisible in the optical (e.g. see Arp 72 and 82 in Figure 2, Arp 202 in Figure 9, and Arp 305 in Figure 14 of \citealt{Smith10}). 

\begin{figure}
\begin{centering}
\plotone{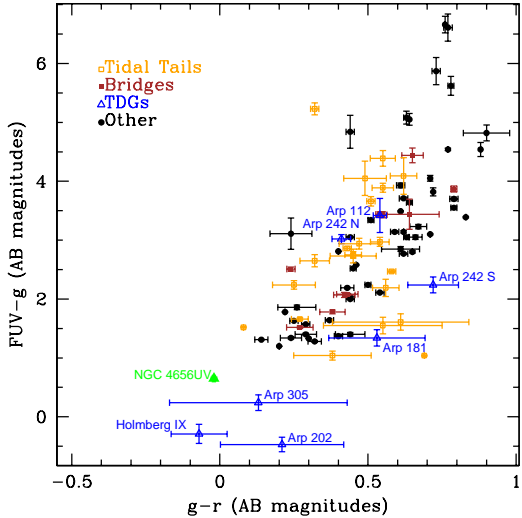}
\caption{FUV-$g$ vs. $g$-$r$ color-color plot for all tidal features in \citet{Smith10}. NGC 4656UV and Holmberg IX are also shown. Features identified by \citet{Smith10} as tidal tails are yellow open boxes, bridges are red filled boxes, TDG candidates are displayed as blue open triangles (with the exception of NGC~4656UV, which is a filled green triangle), and all other objects are black circles.}
\label{smithccplot}
\end{centering}
\end{figure}

In Figure \ref{smithccplot}, we present the $g$-$r$ vs. FUV-$g$ colors of all \citet{Smith10} features with SDSS data compared to NGC~4656UV. We include Holmberg IX, a similarly young TDG near M81, for whose designation is stronger than most candidate TDGs due to resolved stellar population studies \citep{Sabbi08}. Since no published SDSS photometry of Holmberg IX exists in the literature, we photometered Holmberg~IX in SDSS $g$- and $r$-bands, and obtained AB magnitudes of $15.18\pm0.08$ in $g$ and $15.25\pm0.05$ in $r$. We applied Galactic extinction corrections based on values given in NED\footnote{The NASA/IPAC Extragalactic Database (NED) is operated by the Jet Propulsion Laboratory, California Institute of Technology, under contract with the National Aeronautics and Space Administration.}, and error bars in Figure \ref{smithccplot} include the uncertainty in transforming the corrections into $g$ and $r$ magnitudes.

Even among tidal features that exhibit properties of very young stellar populations, NGC~4656UV is one of the bluest objects. A visual inspection shows that the other very blue tidal dwarf candidates --- the TDGs in Arp 202 and Arp 305, as well as Holmberg IX --- are similar in appearance to NGC~4656UV. They all appear clearly in GALEX, but are nearly invisible in SDSS. The apparent $\sim$1 magnitude gap between NGC~4656UV, Holmberg IX, Arp 305, Arp 202, and the rest of the \citet{Smith10} sample is intriguing, although without a larger sample it is impossible to tell if it is significant. (Note that NGC~4656's tail would bridge this gap if plotted on Figure \ref{smithccplot}).  The gap between the bluest and reddest TDG candidates may indicate that the red TDGs are a more evolved form of NGC~4656UV-like objects, or that there are two separate populations of TDGs.

As the only other known candidate TDG within 10 Mpc, comparisons between NGC~4656UV and Holmberg~IX are especially interesting. From resolved stellar population studies, Holmberg IX boasts a very young stellar population, directly comparable to the estimate of $\sim270$ Myr for NGC~4656 \citep{Sabbi08}, and unlike many of the ``classical'' TDG candidates \citep{Duc00}, neither are located at the end of an obvious, long (several 10s of kpc) tidal tail.  Although \citet{Sabbi08} could not completely rule out the presence of a skeleton red giant branch population of stars, Holmberg~IX is one of the most well-established TDG candidates.  The 10 Mpc threshold is roughly the limiting distance for which HST can do resolved stellar population studies, making NGC~4656UV one of the few known tidal dwarf galaxy systems eligible for follow-up HST observations.

\subsection{Other possible origin scenarios}

We cannot rule out the possibility that NGC~4656UV is a pre-existing, low surface-brightness, gas-rich dwarf galaxy that has been highly rejuvenated by its interaction with NGC~4656, or by the interaction between group members NGC~4656 and NGC~4631.  This scenario can be tested attempting to detect an old stellar population's red giant stars through resolved stellar population studies (similar to that done by \citet{Sabbi08} on Holmberg IX), by determining NGC~4656UV's molecular gas content \citep{Braine01}, or by conclusively determining its metallicity through, for example, oxygen abundances in \hii\ regions \citep{Duc94,Weilbacher00}. A confirmation of the low metallicity estimated from the evolutionary synthesis modeling could be an argument against NGC~4656UV forming strictly from tidal debris of its parent galaxy.

We consider that NGC~4656UV is actually a projected tidal tail, viewed at an angle such that it appears to be a real overdensity of stars and gas. We count this scenario as unlikely based on the velocity gradient of NGC~4656/NGC~4656UV shown in Figure \ref{major}.  \citet{Bournaud04} use simulations to show the signature of long tidal tails viewed edge-on in projection is revealed by a change in the velocity gradient along the line of sight: the velocity appears to rise with radius to a point before falling back towards the systemic velocity of the parent galaxy.  Although we cannot completely rule out this scenario, the p-v diagram (Figure \ref{major}) shows that the velocity across the major axis of the TDG candidate may be continuing to increase with distance from the center of NGC~4656.

Finally, we cannot rule out that NGC~4656 is a UV-bright stellar complex in the outer disk of NGC~4656.  These regions, uncovered by GALEX observations, are proving to be common in low density environments in the extreme outer disks of spiral galaxies \citep{Thilker05,Thilker07}.  They reside beyond the classic optical star-forming disk, and trace the extended spiral arm structure revealed by \hi\ observations.  These stellar complexes are found in the peaks of the \hi\ distribution, however, the total gas surface density is still below the classic star formation threshold traced by the H$\alpha$ and $\Sigma_{gas}$ relation \citep{Martin01}.  In fact, these regions of star formation frequently do not have associated H$\alpha$ emission.  Similarly, there is no obvious H$\alpha$ emission associated with NGC~4656UV (see Figure 1 of \citealt{Donahue95}). However, the absence of H$\alpha$ could be due to the fact that young stars from a burst of star formation are visible for a longer period in the UV.

\subsection{Importance of UV data for TDG detection}
UV-bright TDGs like NGC~4656UV demonstrate that, in order to make a complete census of both tidal debris around interacting systems and the star formation density in the local Universe, the ultraviolet part of the spectrum cannot be ignored. A UV-selected sample is needed to further our understanding of these enigmatic dwarfs which are too faint in the SDSS wavebands to be easily detected in an optical survey. Four of the 10 TDG candidates in Figure \ref{smithccplot} with UV and optical photometry lie in a different area of color-color space than all other tidal features, with FUV$-g \approx g-r \approx 0$. While it is unclear exactly what fraction of all TDG candidates are this blue, the relatively large number of UV-bright TDG candidates in this small sample imply that this population is not negligible. TDGs may be common: \citet{Hunsberger96} estimate that up to half of all compact group dwarfs are TDGs, while \citet{Duc99} indicate that a significant fraction of dwarfs in clusters could be recycled objects. Having an accurate census of the byproducts of galaxy interactions is critical to understanding the evolution of groups and clusters, but without an unbiased UV survey a significant fraction of these tidal remnants will remain hidden from view. 

\section{Conclusion}
\label{conclusion}
We have identified a candidate tidal dwarf galaxy around NGC~4656 in GALEX near- and far-UV images. A significant amount of multi-wavelength archival data covering this object exists, including images from the FUV to 160 $\mu$m and an \hi\ data cube from the VLA. 
NGC~4656UV is chiefly characterized by an asymmetric, spatially distinct \hi\ envelope, an extremely blue SED, and a lack of infrared emission associated with dust.  The TDG candidate is currently undergoing a significant burst of star formation, making it extremely bright in the UV. We are unable to detect any emission in survey and archival data further red than 3.6 $\mu$m, suggesting that there is little to no detectable large grain or PAH dust emission in this system, although deeper observations are warranted.

The \hi\ morphology of the entire NGC~4656 system is extremely disturbed featuring warps as well as counterrotating and extraplanar gas. We attempted to fit a rotation curve to NGC~4656 and found that this galaxy does not have the regular rotation expected from normal spiral galaxies. The HI gas is extends high above the disk of NGC~4656 and has a large velocity dispersion, suggesting a previous gravitational encounter and a strong recent burst of star formation. 

We used the \hi\ velocity gradient across NGC~4656UV to make an estimate of the dynamical mass and the virial ratio, finding that for a set of simple assumptions the TDG may be either bound or unbound and that it has a low ratio of dark-to-luminous matter. Both of these results indicate a system that has been formed via galaxy interactions and not from its own local matter over density.  By fitting evolutionary synthesis models from GALEV to the spectral energy distribution of the tidal dwarf galaxy candidate, we find that NGC~4656UV's ongoing starburst is best represented by an extremely low-metallicity model with an age of $\sim$270 Myr and no dust extinction. 
Supported by a simple dynamical argument, we propose that the gas out of which NGC~4656UV is forming stars was stripped from its parent galaxy, NGC~4656, in a tidal interaction with fellow group member NGC~4631 approximately 200-300 Myr ago. 

When compared against other tidal features of interacting galaxies, we find that NGC~4656UV is bluer than tidal tails and bridges in a relatively large sample of tidal debris \citep{Smith10} and it lies in a distinct region of the plot with three other TDG candidates. It is unclear whether these blue candidate TDGs represent the youngest tidal dwarfs or are a distinct population from TDG candidates with redder colors.

NGC~4656UV is an intriguing example of star formation in a low density environment, revealed through observations spanning almost the entire electromagnetic spectrum. Regardless of whether they are primorial in origin or formed from second-hand material, diffuse, gas rich systems like NGC~4656UV may be important sites of ongoing star formation in the nearby Universe.

\acknowledgements{
The authors are deeply in debt to J.~S.~Gallagher for many extremely useful discussions and advice. We thank M.~A.~Bershady and E~M..~Wilcots for comments and suggestions. We are also grateful to B.~Burkhart, N.~M.~Gosnell, and C.~M.~Wood for their deep analysis of this work. We thank an anonymous referee for many comments and suggestions that greatly improved the scope and quality of this work.

Support for this project comes from NSF grants AST-0804576 and AST-1009471.

The National Radio Astronomy Observatory is a facility of the National Science Foundation operated under cooperative agreement by Associated Universities, Inc. 

This work is based in part on observations made with the Spitzer Space Telescope, which is operated by the Jet Propulsion Laboratory, California Institute of Technology under a contract with NASA.

This publication makes use of data products from the Two Micron All Sky Survey, which is a joint project of the University of Massachusetts and the Infrared Processing and Analysis Center/California Institute of Technology, funded by the National Aeronautics and Space Administration and the National Science Foundation.

Funding for the SDSS and SDSS-II has been provided by the Alfred P. Sloan Foundation, the Participating Institutions, the National Science Foundation, the U.S. Department of Energy, the National Aeronautics and Space Administration, the Japanese Monbukagakusho, the Max Planck Society, and the Higher Education Funding Council for England. The SDSS Web Site is http://www.sdss.org/.

The SDSS is managed by the Astrophysical Research Consortium for the Participating Institutions. The Participating Institutions are the American Museum of Natural History, Astrophysical Institute Potsdam, University of Basel, University of Cambridge, Case Western Reserve University, University of Chicago, Drexel University, Fermilab, the Institute for Advanced Study, the Japan Participation Group, Johns Hopkins University, the Joint Institute for Nuclear Astrophysics, the Kavli Institute for Particle Astrophysics and Cosmology, the Korean Scientist Group, the Chinese Academy of Sciences (LAMOST), Los Alamos National Laboratory, the Max-Planck-Institute for Astronomy (MPIA), the Max-Planck-Institute for Astrophysics (MPA), New Mexico State University, Ohio State University, University of Pittsburgh, University of Portsmouth, Princeton University, the United States Naval Observatory, and the University of Washington.

This research made use of Montage, funded by the National Aeronautics and Space Administration's Earth Science Technology Office, Computational Technnologies Project, under Cooperative Agreement Number NCC5-626 between NASA and the California Institute of Technology. The code is maintained by the NASA/IPAC Infrared Science Archive.}

\end{document}